# Electrical energy prediction study case based on neural networks

**Cristian Vasar, Octavian Prostean, Ioan Filip, Iosif Szeidert**

Department of Control System Engineering,

"Politehnica" University of Timisoara, Faculty of Automation and Computers

1900 Timisoara, Bv. V. Parvan, No.2, ROMANIA, vcristi@aut.utt.ro

*Abstract: This paper presents some considerations regarding the prediction of the electrical energy consumption. It is well known that the central element of a microeconomic analysis is represented by the economical agents actions, actions that follow their own interest such as: the consumer – maximization of his satisfaction, the producer – maximization of his profit. The study case is focused on the prediction of the sold energy in Banat region. The goal of this study case is to optimize the electrical energy quantity purchased from the producer by the energy distributor in Banat region. The prediction is based on neural networks. There are used feed-forward and Elman type neural networks. In order to enhance the prediction accuracy there have been used both linear and nonlinear preprocessing units. The aspects considered in this paper can be extrapolated in any general case of prediction-based application, not only in the already stated case of electrical energy.*

## 1 Introduction

The main goal of this paper is to obtain an accurate sold electrical energy prognosis by the main distributor in Banat region. This study is required for an optimization of the purchase costs and implicit of the profit.

In a general case, through production costs is understood the monetary expression of all expenses supported by an economical agent in order to produce, deliver and sale its products or services.

A classification of the costs can be the following:
- transaction costs – represents by expenses that assure the market's functionality;
- information costs – represented by transaction's variables removal (assurance fees, publicity, etc);

- book-keeping costs – defines as the monetary expression of resources used in the production and sale process of a certain quantity of economic goods.

In an economy with limited resources, the problem of alternatives use has a major importance so that the cost is only an alternative measure, an indicator that allows choosing the optimal solution of resource usage.

The production costs can be predicted, traced and analyzed through many criterions, but certainly the most used is one that refers to the analyzed time period. So, there can be distinguished short-term costs and long-term costs. However, there must be noticed that, in economical context, the short/long term is not related to a certain time period. The short/long term is not measured in months or years. Establishing the difference between the mentioned periods requires an effective economical analysis in concordance with a precise situation.

The study case is focused on electrical energy consumption prediction. There is known the tact that if the energetic system operates near the limit of his capacity, the any additional energy request leads to an increase of the undelivered energy, respectively to the increase of the costs due to undelivered energy.

The electric energy request shows significant fluctuations in concordance to seasons and even with the daytime and this is the main reason, which leads to the idea that the additional requests should by satisfied by different production units.

Therefore, results the necessity of using a scientific method that allows an easy adaptation on current energetic power demand. [1]

On the basis, of the analysis of differences recorded between the purchased and sold energy by the Banat region's distributor, the energy losses can be predicted.

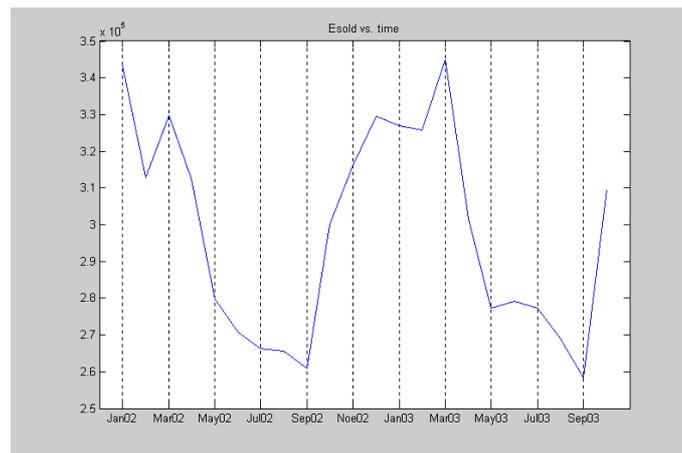

Figure 1. The evolution of the sold energy in Banat area

In Figure 1 the sold energy evolution can be noticed for the Banat region in the last two years. The time (measured in month from Jan 2002 to Oct 2003) is represented on the abscise axis, and the sold energy (measured in MWh) corresponds to the ordinate.

Among the variety of prediction methods, neural networks have been widely used. They can approximate any continuous multivariate function to a desired degree of accuracy, provided that there are a sufficient number of hidden neurons [2].[3]

In this paper there are compared the results of the prediction based on a classic neural network versus a special designed neural network which presents a nonlinear preprocessing unit.

## 2  Prediction structures

There were used two kind of neural networks: feed-forward neural network and Elman network. They were trained by using the gradient descent with momentum backpropagation. The best network structure had two nonlinear layers with hyperbolic tangent sigmoid transfer function and one output linear layer with linear transfer function.

The training method involves that each variable is adjusted according to gradient descent with momentum rule, [4]

$$dX = mc * dX_{prev} + lr * (1 - mc) * \frac{dperf}{dX} \qquad (1)$$

where   dXprev is the previous change to the weight or bias.
lr is the learning rate
mc is the momentum constant.

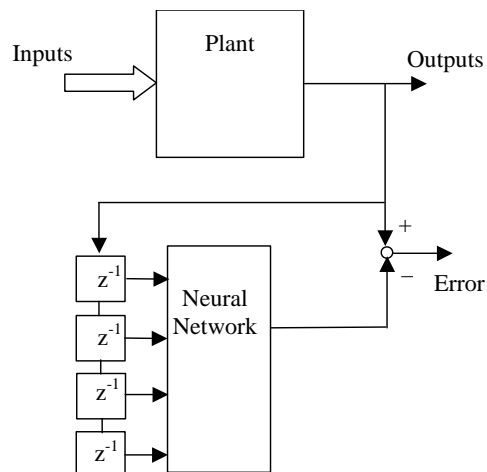

Figure 2. Series-parallel prediction structure

The prediction structure used for training the neural network is depicted in Figure 2. In the series-parallel prediction structure, the neural network receive as inputs the past outputs of the real system [5]

To improve the accuracy of prediction the network inputs pass through a linear preprocessing block which limits the variation of the neural network inputs in the range of [0,1]. This is achieved by a simple mathematical multiplying operation performed over the measured inputs. Also the network output was forced to be in the same range, so a postprocessing unit is required.

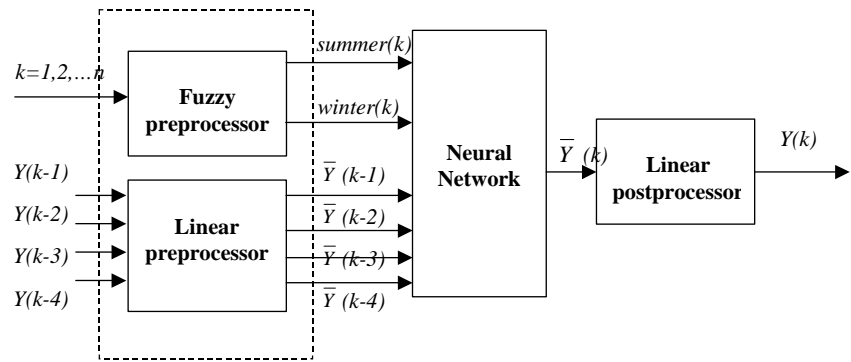

Figure 4. Neural Network with pre/post-processing units

In order to take into consideration of other factors, such as outdoor temperatures, daytime, vacations periods etc, there has to be considered a supplementary set of inputs to the network. All these perturbations are correlated with the four seasons. To minimize the number of network inputs (and consequently the neural network complexity), this supplementary set was reduced to only 2 supplementary inputs, referred generically as "summer" and "winter". This was accomplished by using a fuzzy preprocessor unit. The obtained neural structure is depicted in Figure 4.

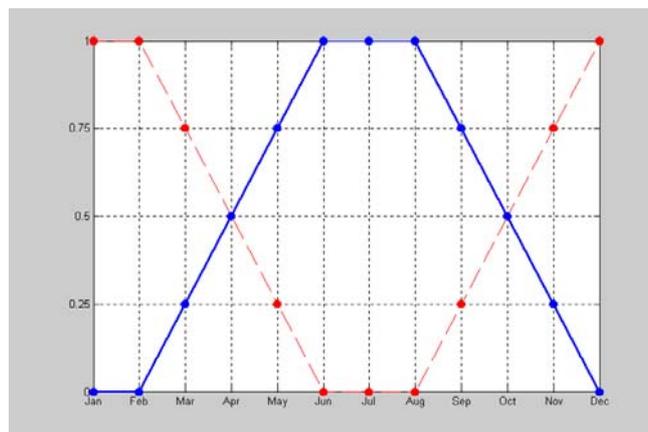

Figure 5. Preprocessor output functions

The input of the fuzzy preprocessor is the prediction month, and the outputs of this represent the membership degree of the current month to the two "seasons", which are referred "winter" and "summer". The preprocessor characteristics are depicted in Figure 5. The dotted line represents "winter" and the solid line is for "summer".

## 3   Simulation studies

There have been created two neural network categories. One consisted of different sized feedforward networks and the other consisted of different sized Elman recurrent networks. The results showed the same trend of prediction error evolution for the entire category, and also proved, in this particular case, that using large networks does not increase the accuracy of prediction.

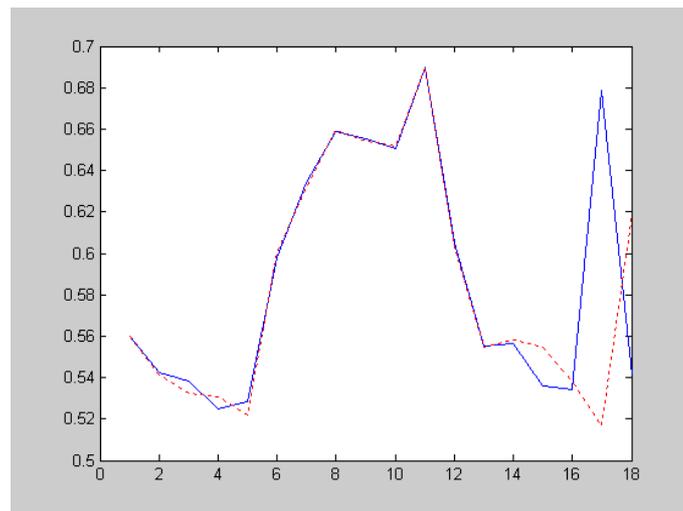

Figure 6. Predicted Sold energy evolution (without "seasons")

The first presented case is the one of the prediction performed by using the series-parallel prediction structure depicted in Figure 2, and using a feedforward network without fuzzy-preprocessing is presented in Figure 6. The first 14 months were used as inputs for the training process, and the last 4 months were predicted. The predicted values correspond to the time interval between May 2002 and October 2003. The real measured values for the sold energy are plotted with dotted line while the predicted valued are depicted with solid line.

As it can be noticed, the outputs presented in Figure 6 and 7 are the scaled ones, and the real predicted sold energy can be obtained by applying this values to the linear postprocessor (which means multiplying by $5*10^5$). The time (measured in month from May 2002 to Oct 2003) is represented on the abscise axis.

In the second case, the same situation was simulated, but the fuzzy preprocessing unit was added to the neural network as depicted in Figure4. The accuracy of prediction increased considerably as it can be noticed in Figure 7. The time (measured in month from May 2002 to Oct 2003) is represented on the abscise axis, and on the ordinate there is represented the scaled sold energy.

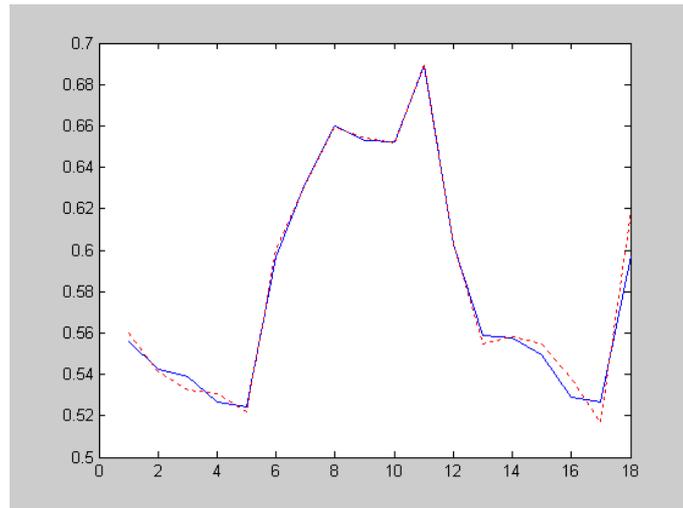

Figure 7. Predicted Sold energy evolution (with "seasons")

In order to obtain predictions for more than 1 step ahead, the trained network, whose behavior was presented in Figure 7, was included into a multi-step prediction structure as the one depicted in Figure 8.

In this situation, the predicted values for the sold energy are obtained by using already predicted values. [6] The predicted results were more than the expectations.

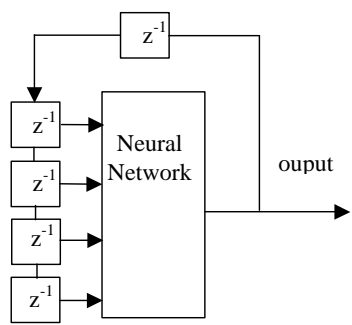

Figure 8. Multi-step prediction structure

The initial inputs for the neural network were the first four measured sold energy values, corresponding to the time interval (Jan-Apr 2002) and the "*summer*" and "*winter*" input corresponding for May 2002. After this point the network did not received any measured sold energy values, and the prediction relies only on already predicted values.

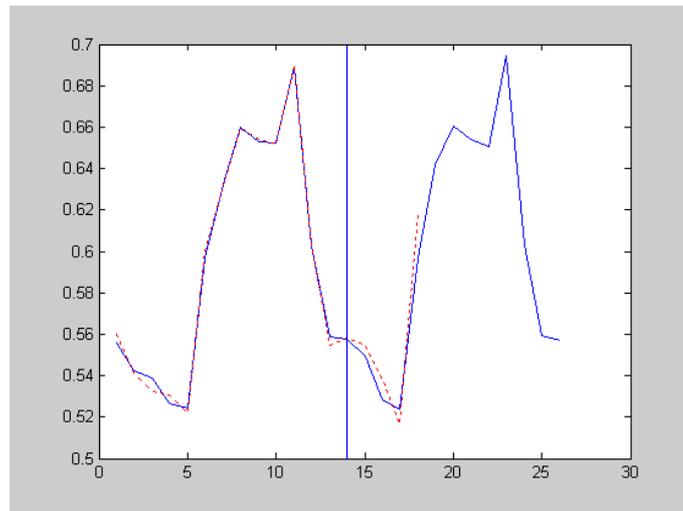

Figure 9. Multi-step simulation for sold energy

The simulation result is depicted in Figure 9. The time (measured in month from May 2002 to Jul 2004) is represented on the abscise axis, and on the ordinate there is represented the scaled sold energy. The first 14 predicted values are closed to the real ones (plotted with dotted line), and this was expectable since they were the source of training, but the network was able to predict further, beyond July 2003.

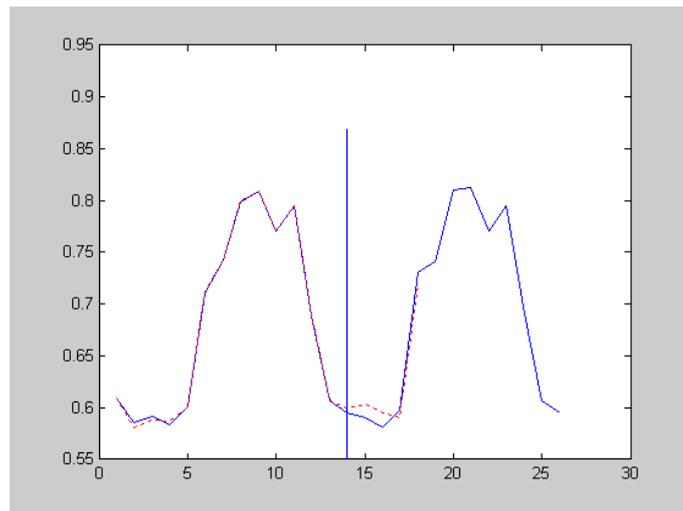

Figure 10. Multi-step simulation for purchased energy

The same experiments were conducted over purchased energy, and the results were quite similar. The purchased energy is depicted in Figure 10. The first 14 months were use as inputs in the training process. The multi-step predicted purchased energy is depicted in Figure 10. The time (measured in month from May 2002 to Jul 2004) is represented on the abscise axis, and on the ordinate there is represented the scaled purchased energy.

**Conclusions**

From the simulations presented in this paper it can be noticed the prediction performance improvement by using the presented nonlinear preprocessor. This unit can very easily implement specific situations as "early/late winter", "early/late summer", by modifying the characteristics presented in Figure 5, in order to fine-tune the prediction (even without network retraining).
In order to persuade this research, the authors have prompted several simulations on selection of proper neural network structures, parameters and optimized length of necessary data.